# COMPARATIVE STUDY OF QOS PARAMETERS OF SIP PROTOCOL IN 802.11A AND 802.11B NETWORK

SUTANU GHOSH

Faculty of Electronics and Comm. Engg,
Dr. Sudhir Chandra Sur Degree Engg College.
sutanu99@gmail.com

*ABSTRACT*

*Present day, the internet telephony growth is much faster than previous. Now we are familiar with digitized packet of voice stream. So, we have required VOIP communication. SIP is one type of VOIP protocol. This one has a SIP proxy. There have one of the important communication environment Wireless LAN (WLAN). WLAN have different radio link standard. Here I am comparing SIP protocol in two radio link standard 802.11a and 802.11b environment. The first one have maximum transmission rate of 54Mbps and second one have maximum transmission rate of 11Mbps. In this paper I want to show the results in a comparative plot. These comparisons include server /client throughput, packet drops, end to end delay etc.*

## 1. WLAN NETWORK

IEEE standard 802.11 a/b/g WLAN architecture consist of one or more BSS service which is called Access point and client devices.

WLAN provide high bandwidth capability. Data rate for 802.11b is 11 mbps where as 802.11 a support data rate 54 mbps. Wireless users can access real time and Internet services virtually anytime anywhere. This services are mainly available in the hot spot like campus, hotel etc. The deployment cost is very low.

## 2. VOIP NETWORK

It defines a process to carry voice calls over the Internet Protocol (IP) network, which includes the digitization and packetized form of the voice streams. A VoIP system consists of an encoder-decoder pair and an IP transport network. The choice of the system vocoder is important because it has to fit the particularities of the transport network (loss and delay).

**Working Procedure:**

Here the voice stream is broken down into small packets, compressed and sent toward their final destination by various routes, depending upon the most efficient paths in a given network congestion. At the other end packets are reassembled, decompressed and converted back into a voice stream by various hardware and software.

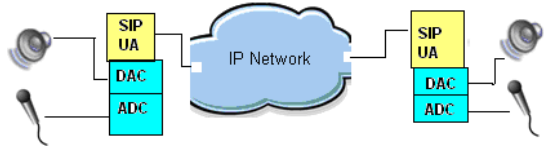

Figure 1: VoIP network

**Advantages of this Network :**

1. VoIP is digital.
2. It can provide the features and services those are not available with a traditional telephony network.
3. There have no call tolls as it uses our internet connection. We can talk with anyone for as long as we want over the internet connection.
4. Voice is converted into the data, transported on the data network and negates the need for a voice network at all.
5. VoIP accept and encircle an open architecture to provide the flexibility.
6. As soon as we log on to any VoIP capable device.

**Voip Packet format :**

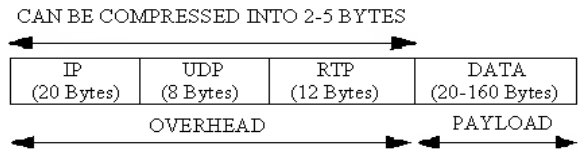

Figure2 : Packet format of VOIP(In bytes)

**RTP Header**: RTP is the real time transport protocol, used for audio or video stream transfer. This parameter is one the main issue for the VOIP application. RTP allows the samples to be reconstructed in the correct sequence and provides a mechanism for the measurement of delay and jitter. The length of RTP header is 12 bytes.

**UDP Header**: UDP is considered as best effort service to add 8 octets and routes the data to the correct destination port. It is a connectionless protocol and does not provide any sequence information or guarantee of delivery.

**IP Header**: IP adds 20 bytes and is responsible for delivering the data to the destination host. It is connectionless and does not guarantee delivery and sequence ordering.

Table 1. Packets loss impact on voice quality

| Percentage of Losses | Voice quality |
|---|---|
| Lower than 5% | Good |
| Higher than 5% | Poor |

## 3. SIP PROTOCOL

Session Initiation Protocol (SIP) is a special type of signaling protocol that controls the initiation, modification and termination of interactive multimedia sessions. The multimedia sessions could be as diverse as audio or video calls among two or more parties or subscriber, chat sessions or game sessions. SIP extensions can be defined for instant messaging, presence and event notifications. SIP is a text-based protocol that is similar to HTTP and Simple Mail Transfer Protocol (SMTP).

**SIP Entities:**

**3.1 User Agent (UA):** The UA is the end point in a SIP connection. IP/Ethernet or IP/WLAN handsets or soft phones are the typical UAs. A gateway that connects the path between packet telephony and a circuit telephony network would also be a UA. This gateway is the main translator in between circuit switch and packet switch network.

**3.2 Registrar:** This is a server that keeps track of an user for the information of the location in the network. This is the main element that allows for the presence tracking or the ability to recognize where a party can be reached and with which modes of communication.

**3.3 Redirect Server:** A server that informs the devices when they must contact different locations to perform a function.

**3.4 Proxy Server:** The primary entity involved in relay signaling messages between end users in a call.

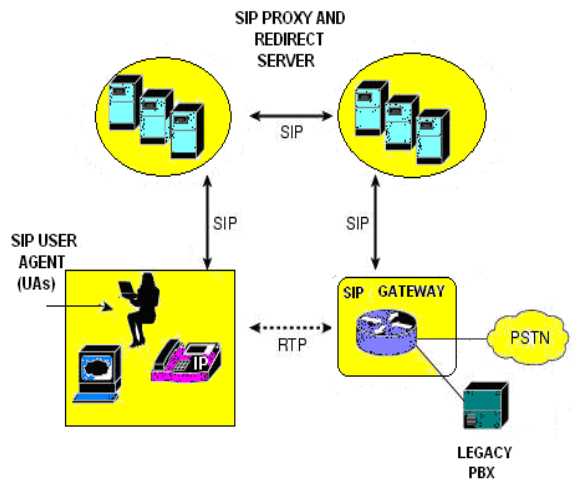

Figure 3 **:** SIP architecture block diagram

**SIP Signaling Messages** :

There have basic five signaling messages are associated with the SIP protocol. Those are –

• **REGISTER:** This message is used to register a User Agent with a SIP Proxy Server.
• **INVITE:** This message is used to initiate a SIP session or call.
• **BYE:** The command used to terminate a SIP session or call.

•**CANCEL:** The command used to abandon a connection attempt that has not yet been completed.
• **ACK:** A message used to acknowledge important notifications regarding the call.

## 4. SIMULATION

Main goal of the simulation is to compare the VOIP scenario over 802.11a and 802.11b network. These comparisons help us to improve the present day application parameters to match with the exact target. Following parameters are required to be specified when setting up the simulation scenario -

- Traffic and Status for VOIP Protocol
- Wireless Settings for MAC type and Radio type and Channel type
- Statistics for output statistics and graph and all other result
- Network protocol
- General settings for Simulation time, Packet interval time and Co-ordinate settings
- Wireless subnet settings
- Mobile host mobility settings
- VOIP and other application settings

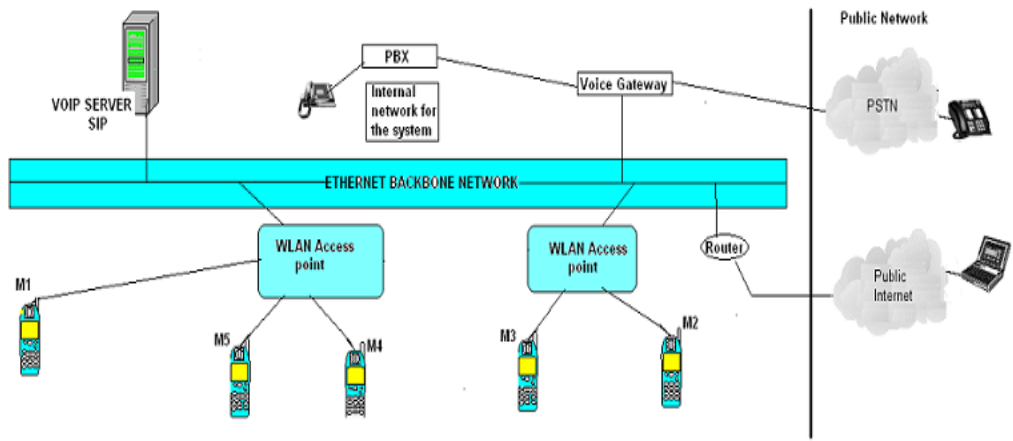

Figure 4: System architecture diagram for a WLAN phone system

## 5. SCENARIO

Here WLAN network have 9 nodes present. Three of them are mobile nodes. There have one extra node 10 (which is used as SIP agent purpose in SIP network and general interface node for H.323 network purpose). Node 1, 2,3,4,5 and node 6,7,8,9 are connected to wireless network1 and wireless network2 respectively. Node 10 is connected to 5 and 6 locally. Here we have taken listening channel as 0100 and listenable channel as 0100 for wireless network 1, listening channel as 0010 and listenable channel as 0010 for wireless network 2 and listening channel as 0001 and listenable channel as 0001 for wireless network 3. We have chosen  host 1, 7 and 8 as  mobile node. Host 1  moves   very slowly through the waypoint 1, 2 and 3 and moves fast at waypoint 4. Host 7 and 8 moves too slow than 1. VOIP connection has chosen in between 4->5, 3->7, 1->9, 2->8, 5->7. Other connection FTP has chosen in between 4->6, 3->8. Another connection CBR have chosen in between 1->9. The total simulation time 134 seconds. The flags are used to represent the path of movement of mobile nodes. The node 10 is used to perform a generalized function for both of the VOIP network(H.323 and SIP).

**Global Parameter Setting -** Three different channels are taken – Two for WLAN network and another for backbone purpose. **39** dBm transmission power is used for all of the 1Mbps, 2Mbps, 6Mbps and 11Mbps. Directional antenna gain is 15dB.

**Channel Frequency, path loss model** –WLAN Network 1, Network 2, Network 3 have the frequency of 2.401GHz, 2.403GHz, 2.402 GHz respectively. Two ray path loss model is used.

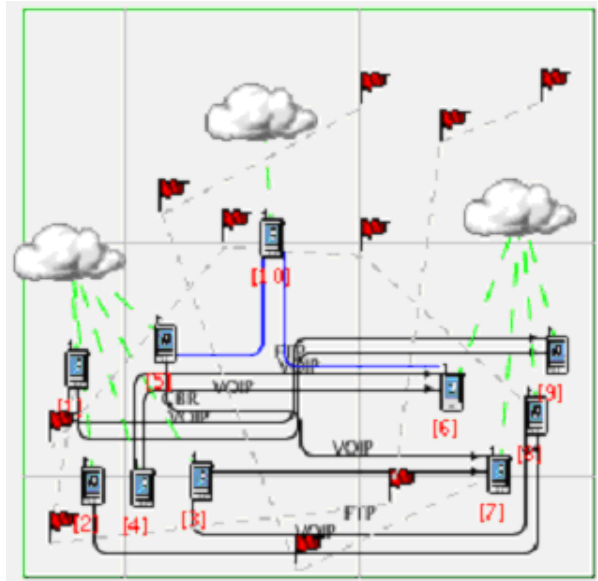

Figure 5 **:** Simulation Scenario

## 6. RESULT

Quality of voice is depending upon the parameters -Throughput, Average Delay and rate of Packet Drop in each scenario have been measured and compared in QualNet 4.5 simulator then it concludes with remarks.

Now, I am showing here some of the graphical plots for the scenario in both of 802.11b and 802.11a network.

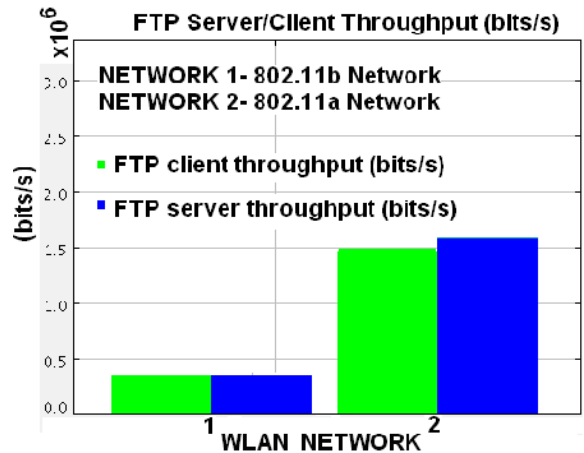

Figure 6 : Plot of FTP server/client throughput

If we see the figure 6 then we can easily conclude that the FTP server throughput of 802.11b is less than 802.11a network. The maximum throughput in the scenario for 802.11a network is 1.6 Mbps. The same scenario have maximum throughput for 802.11b network is 0.4 Mbps.

The comparative study of FTP server throughput and FTP client throughput is based on the node number [4,6] and [3,8] respectively.

In the plot of figure 6 FTP client throughput of 802.11b is less than 802.11a network. The maximum throughput in the scenario for the client in 802.11a network is 1.52 Mbps. The same scenario have maximum throughput for the client in 802.11b network is 0.4 Mbps.

This Figure 6 is the comparative study of FTP with SIP application.

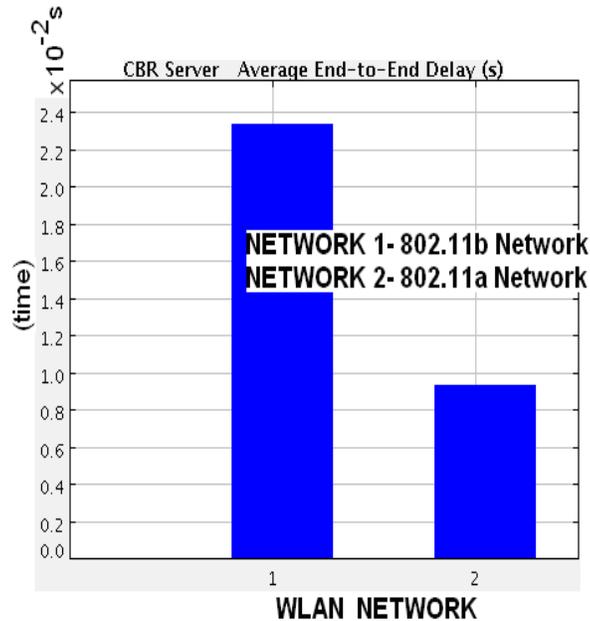

Figure 7: Plot of CBR server average end to end delay

The end to end delay is higher for 802.11b network for CBR server. The amount of delay is 23 ms. It is greater than the delay (9 ms) of 802.11a network. I considered G.711 VOIP traffic in my scenario.

It is very easy to explain that the cause behind this delay and should be the data rate of WLAN network, nothing but else.

CBR server throughput is same for both of the network.

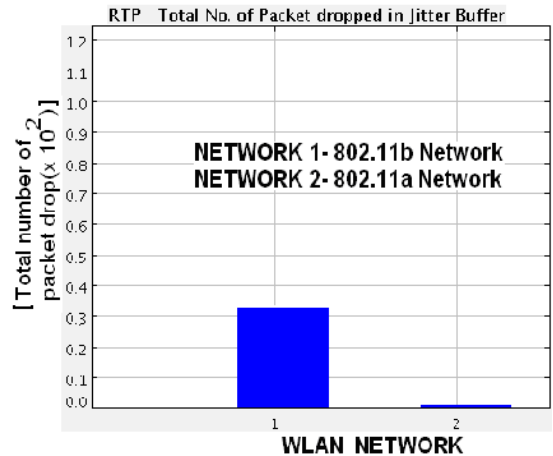

Figure 8 : Plot of total number of packet dropped in jitter buffer

802.11b network have total 32 packets dropped in jitter buffer. Approximately, there have no packet dropped in jitter buffer for 802.11a network. So, according to table-1, it can be proved that the voice quality of 802.11a network is much better than the 802.11b network.

More link frames have been sent for 802.11b network than 802.11a network. But at the receiver same number of frames has been received.

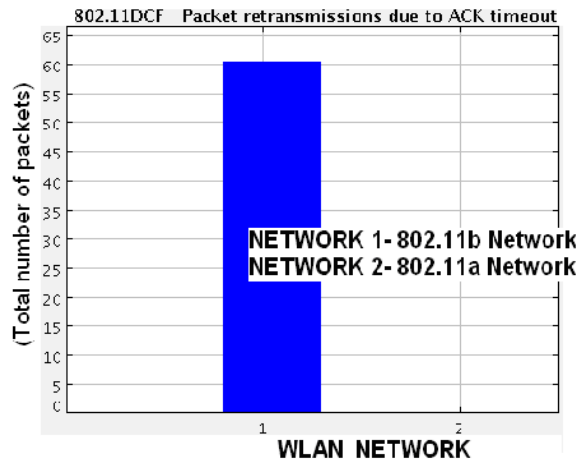

Figure 9 : Plot of 802.11DCF packet retransmission due to ACK timeout

There have required packet retransmission. Because the Acknowledgment time crossed the limiting time. The number of packets retransmission is equal to 61 for 802.11b network and nil for 802.11a network.

Table 2. Comparative study for 802.11a and 802.11b

| Parameter | 802.11a | 802.11b |
| --- | --- | --- |
| Session Initiation time | Minimum 50sec, maximum 175 sec | Minimum 50sec, maximum 175 sec |
| Session establishment time | Minimum 60sec, maximum 125 sec | Minimum 60sec, maximum 125 sec |
| Session End time | All sessions have taken same time to close (135 sec) | All sessions have taken same time to close (135 sec) |
| Talking time | Initiator talking time- Minimum 19 sec Maximum 53 sec  Receiver talking time- Minimum 5 sec Maximum 56 sec | Initiator talking time- Minimum 19 sec Maximum 27 sec  Receiver talking time- Minimum 3 sec Maximum 50 sec |
| Total number of packet sent | Initiator: Minimum 950 Maximum 2550  Receiver: Minimum 200 Maximum 2350 | Initiator: Minimum 600 Maximum 1350  Receiver: Minimum 200 Maximum 2250 |
| Total number of packet Received | Initiator: Minimum 200 Maximum 2350  Receiver : Minimum 950 Maximum 2550 | Initiator: Minimum 200 Maximum 2250  Receiver : Minimum 600 Maximum 1350 |
| RTP session starting time | Minimum : 70 sec Maximum : 135 sec | Minimum : 70 sec Maximum :135 sec |
| Number of RTP packet sent | Minimum : 150 Maximum : 2300 | Minimum : 150 Maximum : 2700 |
| Number of RTP packet received | Minimum : 150 Maximum : 2300 | Minimum : 150 Maximum : 2600 |

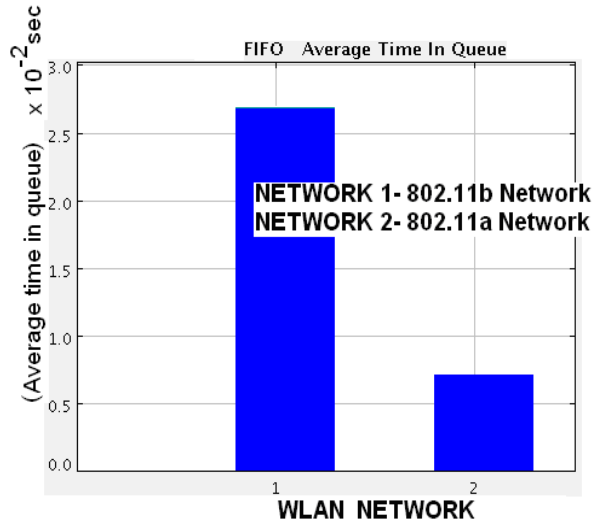

Figure 10 : Plot of FIFO average time in queue

The average time of packets in the queue is 27 ms for 802.11b network and 7.5 ms for 802.11a network.

## 7. CONCLUSION

The main goal of this work was to gain some insight of the relation between the distance on the communications and the quality of voice in different WLAN domain. The measurements inside the busiest environment I observed that the link becomes unstable in the presence of obstacles. My experimental results proved consistent as we observed that 802.11a network link quality and consequently the voice quality much better than 802.11b network. This comparative literature is much more helpful for next future work to measure the voice quality standard of another type VOIP protocol (H.323).

## Author

Sutanu Ghosh has completed his Masters in Mobile Computing and Communication Engineering from Jadavpur University, Kokata, India in 2009. Presently he is an Asst. Prof in Dr Sudhir Chandra Sur Degree Engg College.

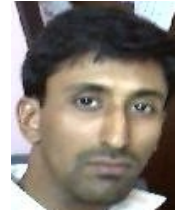